\documentclass[twocolumn]{aastex631}

\newcommand{\kms}{\hbox{km s$^{-1}$}}

\newcommand{\chthreeoh}{\mbox{CH$_3$OH}}
\newcommand{\cIItwosixfive}{\mbox{6$_{1,5}$-5$_{2,4}$E}}
\newcommand{\cIItwofourseven}{\mbox{4$_{2,2}$-5$_{1,5}$A}}
\newcommand{\cIItwofournine}{\mbox{16$_{3,13}$-15$_{4,12}$E}}
\newcommand{\cIItwosixone}{\mbox{2$_{1,1}$-1$_{0,1}$E}}
\newcommand{\cItwofivezero}{\mbox{11$_{0,11}$-10$_{1,10}$A}}
\newcommand{\cItwosixsix}{\mbox{5$_{2,4}$-4$_{1,3}$A}}
\newcommand{\ciizinchenko}{\mbox{14$_{1}$-14$_{0}$A$^{-+}$}}
\newcommand{\thermal}{\mbox{8$_{3,5}$-8$_{2,6}$A}}
\newcommand{\jypbeam}{\mbox{Jy~beam$^{-1}$}}

\begin{document}

\title{Spectral survey of a Hot core with an Eruptive Accretion in S255IR NIRS3 (SHEA):\\
 The discovery of class I and class II millimeter methanol maser transitions}

\author{Giseon Baek}
\affiliation{Department of Physics and Astronomy, Seoul National University, 1 Gwanak-ro, Gwanak-gu, Seoul 08826, Korea}
\affiliation{Research Institute of Basic Sciences, Seoul National University, Seoul 08826, Republic of Korea}

\author{Jeong-Eun Lee} 
\affiliation{Department of Physics and Astronomy, Seoul National University, 1 Gwanak-ro, Gwanak-gu, Seoul 08826, Korea}
\affil{SNU Astronomy Research Center, Seoul National University, 1 Gwanak-ro, Gwanak-gu, Seoul 08826, Republic of Korea}

\author{Neal J. Evans II}
\affiliation{Department of Astronomy, The University of Texas at Austin, 2515 Speedway, Stop C1400, Austin, Texas 78712-1205, USA}

\author{Tomoya Hirota}
\affiliation{Mizusawa VLBI Observatory, National Astronomical Observatory of Japan, Osawa 2-21-1, Mitaka-shi, Tokyo 181-8588, Japan}
\affiliation{Department of Astronomical Sciences, SOKENDAI (The Graduate University for Advanced Studies), Osawa 2-21-1, Mitaka-shi, Tokyo 181-8588, Japan}

\author{Yuri Aikawa}
\affiliation{Department of Astronomy, Graduate School of Science, University of Tokyo, Tokyo 113-0033, Japan}

\author{Ji-hyun Kang}
\affiliation{Korea Astronomy and Space Science Institute, 776 Daedeok-daero, Yuseong, Daejeon 34055, Republic of Korea}

\author{Jungha Kim}
\affiliation{Korea Astronomy and Space Science Institute, 776 Daedeok-daero, Yuseong, Daejeon 34055, Republic of Korea}

\author{Jes K. J\o{}rgensen}
\affiliation{Niels Bohr Institute, University of Copenhagen, Copenhagen K., Denmark}

\begin{abstract}
We report the detection of the millimeter $\chthreeoh$ masers including a new detection of class I (\cItwofivezero) and class II (\cIItwosixfive) maser transitions toward the high-mass protostar S255IR NIRS3 in post-burst phase. 
The $\chthreeoh$ emissions were detected as a mixture of maser and thermal characteristics. We examine the detected transitions using an excitation diagram and LTE model spectra and compare the observed properties with those of thermal lines. Class II $\chthreeoh$ maser transitions showed distinctive intensity and velocity distributions from those of thermal transitions.
Bright distinct emission components in addition to the fragmented and arc-shaped emissions are only detected in class I $\chthreeoh$ maser transitions toward southern and western directions from the protostellar position, implying the presence of the slow outflow shocks. 

\end{abstract}

\section{Introduction} \label{sec:intro}
Recently, there is growing evidence that the accretion process of high-mass young stellar objects (HMYSOs) is a scaled-up version of their low-mass counterparts, which undergo the episodic accretion burst phase interspersed with the long quiescent phase.  
The simulations predict that HMYSOs could obtain up to $\sim$50\% of the final mass by the repetitive accretion bursts interspersed by long quiescent phases \citep{Meyer2017MNRAS.464L..90M,Meyer2019MNRAS.482.5459M}, implying that the burst event is crucial for the formation process of HMYSOs.

S255IR NIRS3 (or S255IR-SMA1) is an HMYSO, located in the S255 region at a distance of 1.78 kpc \citep{Burns2016}, harboring a chemically rich hot core. The accretion-mediated luminosity burst was detected for the first time in HMYSOs. The burst started in mid-2015, recognized by the class II $\chthreeoh$ maser flare \citep{Fujisawa2015ATel.8286....1F}. 
The scattered light through the outflow cavities also showed a brightness increase and extended structure from the central star during the burst.
The luminosity increased from 2.9$\times$10$^{4}$ to 1.6$\times$10$^{5}$ L$_{\odot}$ (a factor of $\sim$5.5) with accretion rate of 5$\times$10$^{-3}$ M$_{\odot}$ yr$^{-1}$ during the accretion burst \citep{Stecklum2016ATel.8732....1S,CarattioGaratti2017}. 

During the burst phase, the peak 6.7 GHz class II $\chthreeoh$ maser luminosity increased and the emitting region extended over the larger radii (500--1000 au), implying that radiation field strength was enhanced \citep{Moscadelli2017,Szymczak2018A&A}. 
A new class II $\chthreeoh$ maser was discovered at 349107.0 MHz $\ciizinchenko$ transition \citep{Zinchenko2017}, without variability within the five-month observation interval. 

Then, the accretion burst lasted for $\sim$2 yrs at submillimeter wavelengths; both continuum flux and 6.7 class II CH$_{3}$OH maser intensity started to decrease, implying the decaying of radiation strength in S255IR NIRS3 \citep{Cesaroni2018,liu18, Szymczak2018A&A,Uchiyama2020PASJ...72....4U}.

The burst events of HMYSOs provide a special opportunity to explore the nature of accretion traced by the $\chthreeoh$ maser as well as search for the new maser transitions (S255IR NIRS3, \citealt{CarattioGaratti2017,Szymczak2018A&A}; NGC6334I-MM1, \citealt{hunter17,MacLeod2019MNRAS.489.3981M}; G358.93-0.03, \citealt{Stecklum2016ATel.8732....1S,Volvach2020MNRAS.494L..59V,Brogan2019ApJ, Chen2020NatAs...4.1170C,Burns2020NatAs...4..506B,Burns2023NatAs.tmp...57B}).
In addition, $\chthreeoh$ maser characteristics at (sub)millimeter wavelengths have started to be investigated due to the development of recent high-resolution interferometry \citep{Zinchenko2017,Brogan2019ApJ, Kim2020,Baek2022ApJ...939...84B, Hirota2022PASJ...74.1234H,Humire2022A&A...663A..33H}, which extends our understanding of the $\chthreeoh$ maser characteristics at shorter wavelengths.

In this paper, we report newly found millimeter class I and class II $\chthreeoh$ maser transitions toward S255IR NIRS3 which recently underwent the accretion burst phase. 
In Section 2, the observation scheme is described. In Section 3, we report the detection of $\chthreeoh$ maser transitions, mixed with thermal emission, with newly discovered transitions as masers. In Section 4, a discussion of previous maser studies in S255IR NIRS3 is presented.

\section{Observation} \label{sec:obs}
We conducted ALMA band 6 observation toward S255IR NIRS3 with two configurations, designed to investigate the detailed behavior along with the infall from envelope to the protostar  (2019.1.00571.S and 2021.1.01056.S: P.I. Giseon Baek). 
Compact and extended observations were conducted in C-5 and C-8 configurations, respectively. The observations were carried out on 2021 April 9, December 2, and 12 for compact configuration, and 2021 October 3, 5, 15, and 16 for extended configuration. 
ALMA pipeline calibration was conducted using CASA 6.2.1 \citep{CASA2022PASP..134k4501C}. 
Imaging was performed using the CASA \textit{tclean} task with a Briggs weighting of 0.5 by combining uv data obtained from C-5 and C-8 configurations. 

The center position of S255IR NIRS3 is (RA, Dec) = (06:12:54.02,+17:59:23.10) in the J2000.0 epoch.
The spectral resolution is 488.267 kHz, corresponding to $\sim$0.609 km s$^{-1}$.
The beam size of the continuum image is 0.045$''\times$0.039$''$. 
As S255IR NIRS3 harbors a line-rich hot core, we utilized STATCONT \citep{SanchezMonge2018}, which statistically estimates the continuum level in each pixel, providing continuum-subtracted line cubes and continuum images. 

\section{Results} \label{sec:results}
As the intrinsic characteristics of the maser are coherent and amplified emission, the maser feature is observationally distinct from thermal emission. The widely-used identification criteria of maser are as follows: compared to the thermal line, the maser line tends to present enhanced intensity (high brightness temperature, T$_{\rm b}$) at certain velocities (narrow line width) in a confined area which satisfies the emerging condition (morphological compactness).

\subsection{1.2 mm continuum map}
Figure \ref{fig_mom_classII}(a) presents an integrated intensity map of $\chthreeoh$ $\thermal$ at 251517.269 MHz (color map) with the 1.2 mm continuum (contour). In the continuum map, the protostellar position is well-constrained. The northern envelope emission of S255IR NIRS3 is broadly resolved into three branches, marked as N-E, N, and N-W.

One new feature detected in our observation is an elongated continuum emission in the NE-SW direction. It is coincident with the outflow direction investigated by near-infrared scattered light, CO and SiO emissions, and 22 and 321 GHz water maser observations \citep{Goddi2007,Zinchenko2015,Burns2016,CarattioGaratti2017,Hirota2021}. 

Right after the burst, a faint N-W elongated radio continuum emission was detected at 1.3 cm (Figure 4 in \citealt{Cesaroni2018}).
The direction of the elongation was close to the outflow axis traced by the H$_{2}$O masers \citep{Goddi2007,Burns2016,Hirota2021}, and thus, the feature was ascribed to the burst-generated radio jet. 
Our 1.2 mm continuum elongation feature on both N-E and S-W directions might also be attributed to the heated dust by the jet ejected from the recent burst. The detailed comparison with previous observations for time variation requires consideration of different responses of observed frequencies at different epochs, which is beyond the scope of this study. 

\subsection{Class II $\chthreeoh$ maser}

\begin{figure*}
  \centering
\includegraphics[width= 0.48\textwidth]{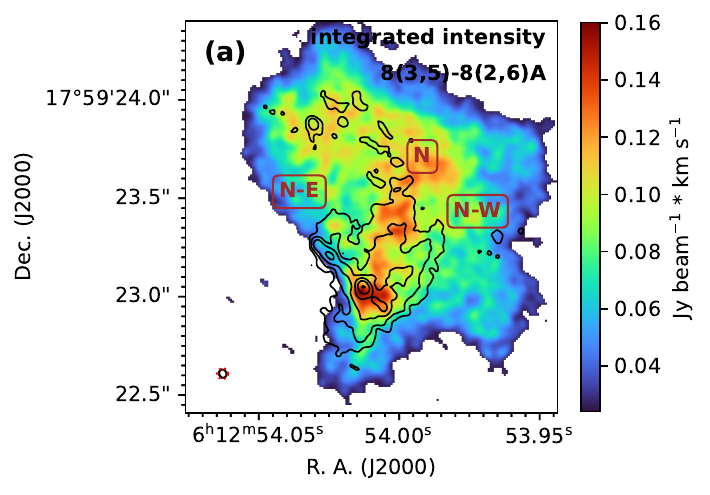}
\includegraphics[trim={1.95cm 0 0 0},clip,width= 0.4\textwidth]{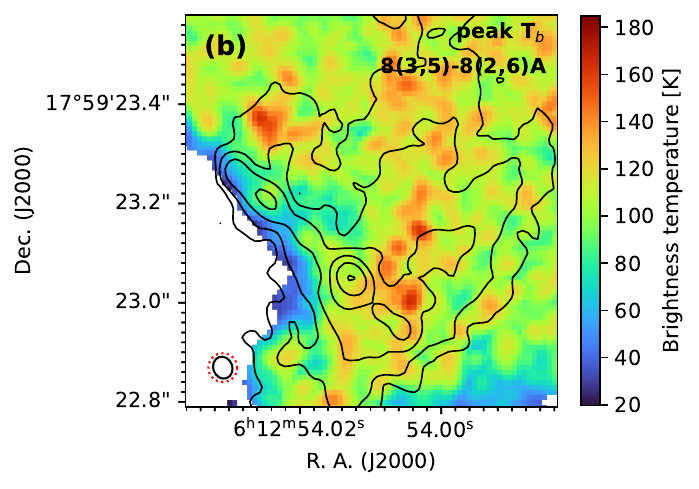}
\includegraphics[width=0.48\textwidth]{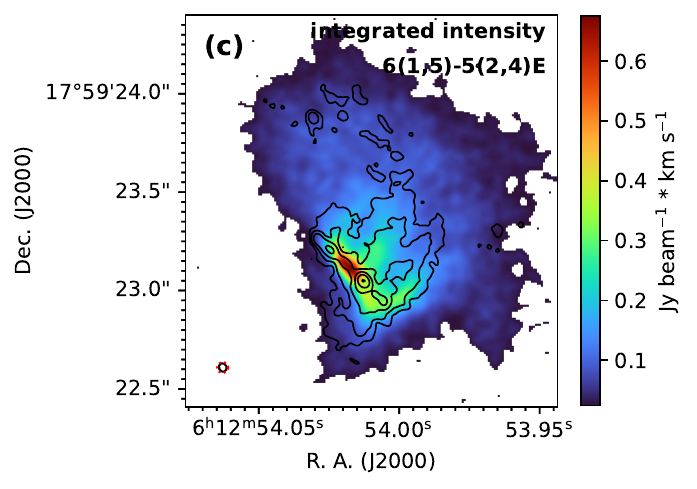}
\includegraphics[trim={1.95cm 0 0 0},clip,width=0.41\textwidth]{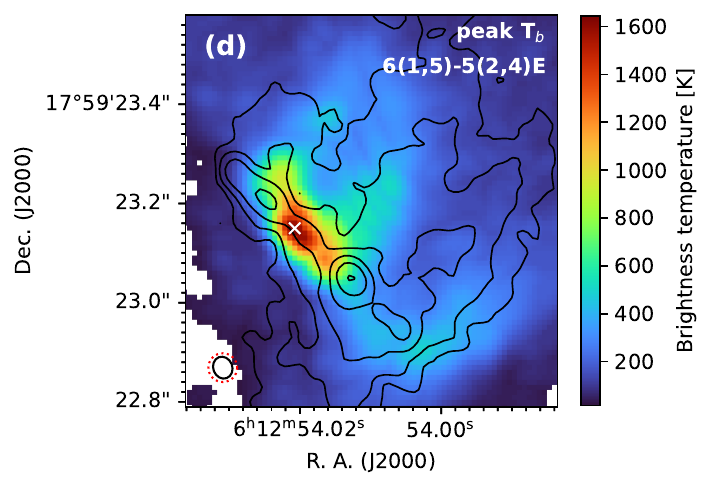}\\
\caption{Upper panel: Integrated intensity (left) and peak T$_{\rm b}$ (right) maps of $\chthreeoh$ $\thermal$ line overlaid with the 1.2 mm continuum map (contour). The intensity at the velocity range of -3.8 and 11.2 km s$^{-1}$ is integrated with $>$ 5$\sigma$ of rms noise. 
The contour levels for the continuum emission are  0.3, 0.7, 1.5, 3.5, 7.9, and 18.0 m$\jypbeam$.
Lower panel: The same maps but for $\chthreeoh$ $\cIItwosixfive$ line. White $\times$ in peak T$_{\rm b}$ map marks the region that the line profiles in Figure \ref{fig:maserline} is extracted.} \label{fig_mom_classII}
\end{figure*}

\begin{figure}
  \centering
\includegraphics[width=0.7\columnwidth]{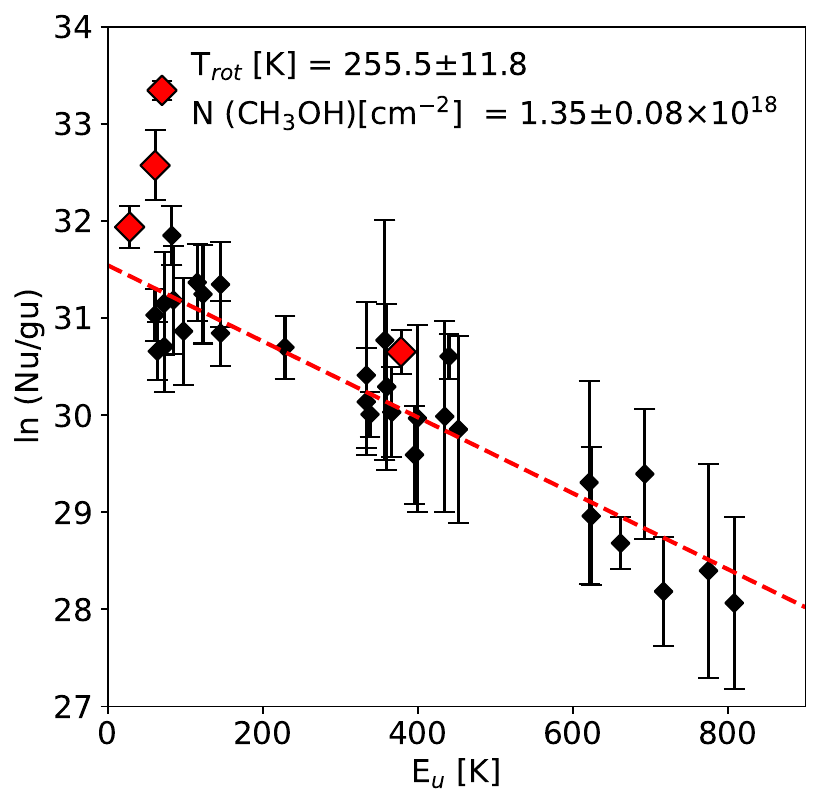}
\hspace{0.5cm}
\includegraphics[width=\columnwidth]{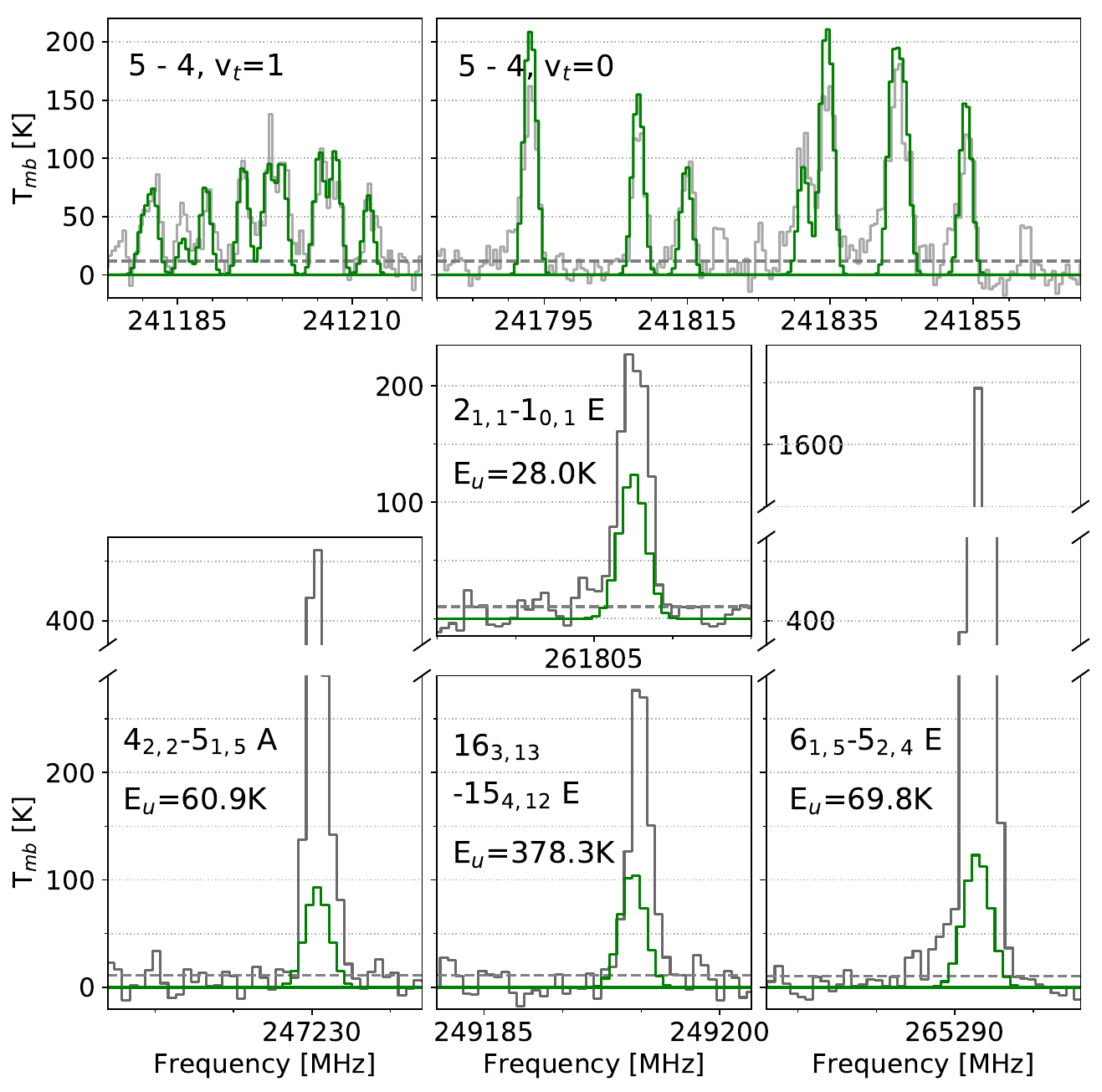}
\caption{Upper panel Excitation diagram of $\chthreeoh$. In a range of 238-275 GHz, emission lines with A$_{ij}$ $<$ -4.25 are adopted. Lower panel: Observed (grey) and model (green) spectra of $\chthreeoh$ in S255IR NIRS3 (x marked position in Figure \ref{fig_mom_classII}d).  } \label{fig:maserline}
\end{figure}

\begin{figure}
  \centering
\includegraphics[trim={0 1.2cm 0 0},clip,width=\columnwidth]{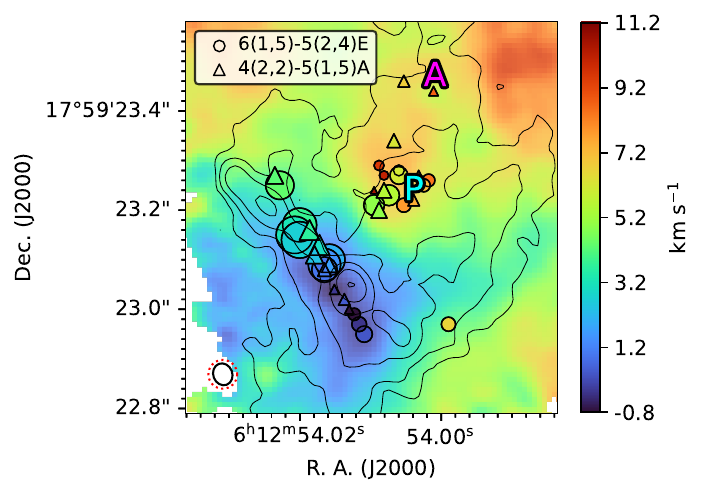}
\includegraphics[trim={0 1.2cm 0 0.2cm},clip,width=\columnwidth]{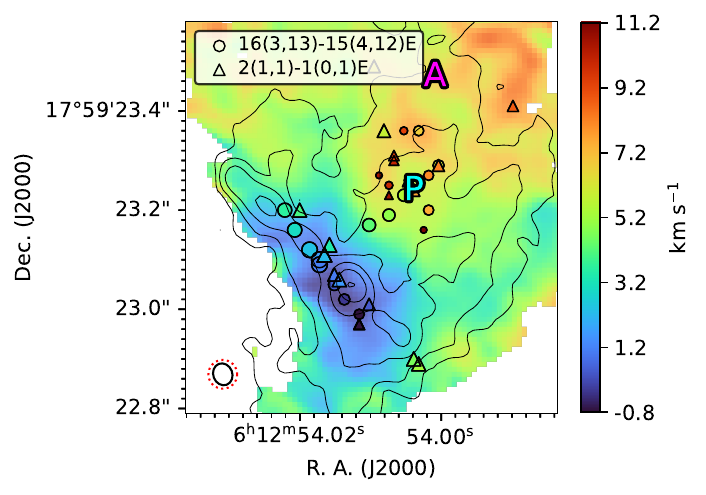}
\includegraphics[trim={0 0 0 0.2cm},clip,width=\columnwidth]{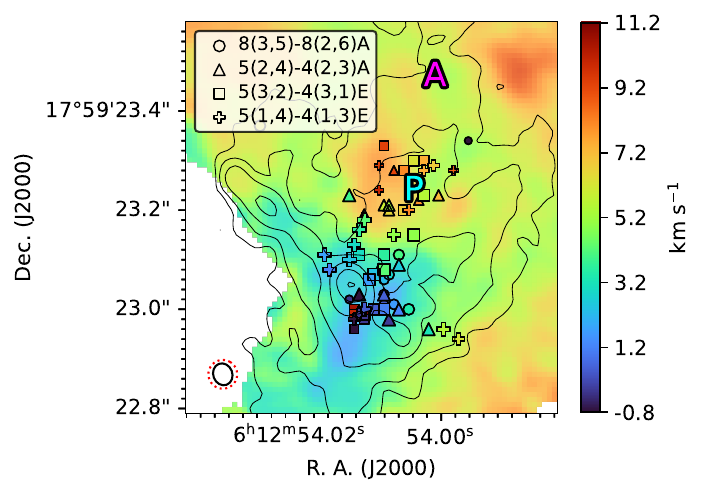}
\caption{
The intensity-weighted velocity maps (colored image) and the velocity distribution of the peak intensity positions (symbol) of $\chthreeoh$ transitions. The contour is 1.2 mm continuum map, the same as Figure \ref{fig_mom_classII}. The intensity-weighted velocity map of the first transition in the legend (the strongest emission) is shown in each panel. The background color image has lightened with $\alpha$=0.8 (0:transparent, 1: opaque). The symbol denotes the location of the emission peak in each channel. The color of the symbol shows the velocity at the peak intensity position, identical to the color image scale. Symbol size is proportional to the intensity. 
P and A denote the 6.7 GHz class II $\chthreeoh$ maser clusters observed before and after the accretion burst of S255IR NIRS3 \citep{Moscadelli2017,Szymczak2018A&A} Top: Transitions that class II $\chthreeoh$ maser is predicted by the model (Table \ref{tab:transitions}). Middle: Class II maser candidate transitions that behave similarly to that presented in the top panel, without model prediction. Bottom: Thermal transitions.}
\label{fig:centroidmap}
\end{figure}

Class II $\chthreeoh$ maser is known to be pumped by the infrared radiation from the heated dust. Observational studies have reported that the class II $\chthreeoh$ maser is a reliable indicator to search for the hot core in the early stage of star formation \citep[e.g,][]{Baek2022ApJ...939...84B} as well as to trace the accretion phenomena \citep{Burns2020NatAs...4..506B,Burns2023NatAs.tmp...57B}.

As an example thermal $\chthreeoh$ emission distribution for comparison, Figure \ref{fig_mom_classII} (a) and (b) present the integrated intensity and peak T$_{\rm b}$ distributions of the $\thermal$ line in the inner envelope of S255IR NIRS3. The range of the peak T$_{\rm b}$ is $\sim$ 100--200 K, often found in $\chthreeoh$ thermal emission in hot cores \citep[e.g.,][]{Hernandez-Hernandez2014}. A relatively high integrated intensity along with the N-branch and near the protostellar position might be attributed to the dense gas infalling to the protostar through the streamlines \citep{Liu2020}. 

In Figure \ref{fig_mom_classII} (c) and (d), the spatial size and T$_{\rm b}$ in the extended envelope region ($<$ $\sim$200 K) of $\cIItwosixfive$ emission are comparable to those of $\thermal$ emission.  
However, the $\cIItwosixfive$ emission is inconsistent with the thermal emission in the N-E branch: the $\thermal$ emission is diminished and has no strong tendency in T$_{\rm b}$ map, while the $\cIItwosixfive$ emission is highly enhanced at the specific position, of which the highest T$_{\rm b}$ is $\sim$1650 K. A slight enhancement is also observed in N and N-W branches. 

In the absence of a detailed model for millimeter class II $\chthreeoh$ maser, the transitions with enhanced T$_{\rm b}$ are examined assuming local thermodynamic equilibrium (LTE). 
Figure \ref{fig:maserline} presents the excitation diagram of $\chthreeoh$. The spectra are extracted at the peak of $\cIItwosixfive$ emission (white $\times$ mark in Figure \ref{fig_mom_classII}(d)). The transitions with log$_{10}$ A$_{ij}$ $<$ -4.25 in 238--275 GHz, excluding maser transitions, are adopted because they are optically thin. Otherwise, the saturated emissions for transitions with log$_{10}$ A$_{ij}$ $>$ -4.25 suffer from the high optical depth, underestimating N$_u$ values,
which prevents reliable temperature and density estimation. 
The elaborate optical depth correction can improve the result.

The excitation diagram renders the $\chthreeoh$ rotation temperature of 255 K and column density of 1.35$\times$10$^{18}$ cm$^{-2}$. We apply them to XCLASS to synthesize the spectrum under the LTE assumption. The bottom panel of Figure \ref{fig:maserline} presents the observed (grey) and modeled (green) spectra. The $\chthreeoh$ 5--4 lines in the ground state (v$_{\rm t}$=0) and torsionally excited (v$_{\rm t}$=1) states are presented as thermal lines. The model with the derived physical parameters reproduces well the observed lines. 

Contrary to the $\chthreeoh$ 5-4 transitions, other $\chthreeoh$ lines in Figure \ref{fig:maserline} are not reproduced by the LTE model: the model predicts T$_{\rm b}$ of $\sim$100 K, but the observed T$_{\rm b}$ peaks are significantly higher than predicted ones. In particular, the $\cIItwosixfive$ line peaks at $\sim$1650 K, which fairly exceeds the thermal temperature commonly found in $\chthreeoh$ rotational lines. The peak T$_{\rm b}$ of the $\cIItwofourseven$, $\cIItwofournine$, and $\cIItwosixone$ transitions are $\sim$500, 300, and 250 K, respectively, which are also unexplained by LTE model spectra. We note that the observed T$_{\rm b}$ is a lower limit because of the insufficient angular and velocity resolutions relative to the actual masing components.

The masing region is spatially resolved in our observation.
The T$_{\rm b}$ of $\cIItwosixfive$ in the extended envelope is comparable to that of $\thermal$, but the enhancement is confined in N-E and N branches (Figure \ref{fig_mom_classII}). In the region, the physical condition may meet the masing condition of the transition. The maser feature in the region could be mixed with the thermal emission through the line of sight and diluted in the insufficient beam size.

The kinematic structure that maser transition is tracing is also distinct from thermal transition.
Figure \ref{fig:centroidmap} shows the intensity-weighted velocity map (color map) with the 1.2 mm continuum map (contour). 
In the intensity-weighted velocity map, two velocity components of the inner envelope are seen (Figure \ref{fig:centroidmap}): the red-shifted one in the north and the blue-shifted one near the protostellar position. The blue-shifted component is more prominent in maser transitions than in thermal transitions.

This difference is shown more effectively in the velocity distribution of the peak intensity positions (symbol). The map is constructed by finding the location of the emission peak in each channel. 
The color of the symbol shows the velocity at the peak intensity position. Symbol size is proportional to the intensity.
In our observation where the maser feature is mixed with the thermal emission, the masing component is efficiently traced by the map as the maser emission intensity exceeds the thermal one.

The velocity distribution of the peak intensity positions of $\cIItwosixfive$ and $\cIItwofourseven$ (top panel of Figure \ref{fig:centroidmap}), and $\cIItwofournine$ and $\cIItwosixone$ (middle panel) show strong blue-shifted components accelerating toward the continuum peak position from both the N-E branch and an opposite S-W side. 
The redshifted component is found in the P position (marked with cyan color). On the other hand, the thermal emission (bottom panel) shows a red- to blue-shifted velocity gradient perpendicular to the NE-SW direction.

The line information of four transitions is listed in Table \ref{tab:transitions}. The brighter two transitions, $\cIItwosixfive$, and $\cIItwofourseven$, are predicted to mase by model \citep{Cragg2005}. The $\cIItwosixfive$ line is detected for the first time as a class II maser line.  
The $\cIItwosixfive$ and $\cIItwofourseven$ transitions have similar upper energy levels (E$_{\rm u}$) of 69.8 and 60.9 K, respectively. Line strengths multiplied by the dipole moment (S$_{\rm ij}\mu^2$) and Einstein A coefficient (A$_{\rm ij}$) are slightly higher for $\cIItwosixfive$ line. Compared to the thermal transition $\thermal$, two transitions have lower S$_{\rm ij}\mu^2$ and A$_{\rm ij}$. However, the peak flux densities (F$_{\rm peak}$) of the two transitions appear to be higher than the thermal one, suggesting that they are excited non-thermally, most likely due to maser activity.

Even without the maser model prediction, $\cIItwofournine$ and $\cIItwosixone$ also show similar behavior with the model-predicted maser transitions, while other thermal transitions show different velocity distributions. 
Their T$_{\rm b}$ are lower than model-predicted transitions although S$_{\rm ij}\mu^2$ and A$_{\rm ij}$ are higher than them. Compared to the model-predicted masers and thermal transition, E$_{\rm u}$ of model-unpredicted transitions are higher and lower as 378.3 and 28 K for $\cIItwofournine$ and $\cIItwosixone$, respectively. It suggests that the physical condition of the inner envelope is at the boundary of the range of their maser emerging condition (e.g., T$_{\rm gas}$), or the region at which meets the emerging condition itself is smaller in the line-of-sight.

\begin{deluxetable*}{llcccccclc}
\tabletypesize{\scriptsize}
\tablecaption{Properties of detected $\chthreeoh$ lines \label{tab:transitions}}
\tablehead{
\colhead{Rest Frequency}  & \colhead{Beam size}& \colhead{Transition} &  \colhead{S$_{\rm ij}\mu^2$(D$^2$)} & \colhead{log$_{10}$(A$_{\rm ij}$)} & \colhead{E$_{\rm u}$}& \colhead{ F$_{\rm peak}$}&\colhead{ $v_{\rm peak}$}&
\colhead{Characteristics\tablenotemark{a}} & \colhead{Model}\\
\colhead{(MHz)}   &\colhead{($''\times''$),P.A.(deg)}&  \colhead{Quantum numbers} & 
\colhead{} & \colhead{}& \colhead{(K)}& \colhead{ (Jy)} & \colhead{ (\kms)} & \colhead{} & \colhead{prediction}
}
\startdata
251517.31 (1.1E-2),& [0.058$\times$0.058],0.0 & 8(3,5)- 8(2, 6) A, v$_{\rm t}$=0 & 	29.21431	& -4.09928 & 133.4 &  2.4\tablenotemark{b} & 5.7\tablenotemark{b} & TH & -- \\
265289.56 (1.2E-2) &[0.058$\times$0.058],0.0&  6(1,5)-5(2,4) E, v$_{\rm t}$=0&  6.18297 &	-4.58771 & 69.8 & 4.3\tablenotemark{b} & 4.2\tablenotemark{b} & TH + class II &  Y\tablenotemark{d}\\
247228.59 (1.3E-2)&[0.058$\times$0.058],0.0 &4(2,2)-5(1,5) A, v$_{\rm t}$=0   & 4.34433	& -4.67315	& 60.9 & 3.0\tablenotemark{b} & 5.5\tablenotemark{b} & TH + class II& Y\tablenotemark{d}\\
249192.84 (1.3E-2) &[0.058$\times$0.058],0.0 & 16(3,13)-15(4,12) E, v$_{\rm t}$=0 & 18.59480	&-4.59564	&378.3 & 1.6\tablenotemark{b} & 5.3\tablenotemark{b} & TH + class II ? & -- \\
261805.68 (6.0E-3)&[0.058$\times$0.058],0.0 &  2(1,1)-1(0,1) E, v$_{\rm t}$=0  & 5.33586	& -4.25395 & 28.0& 3.3\tablenotemark{b} & 5.7\tablenotemark{b} & TH + class II ?& -- \\
250506.85 (1.1E-2)&[0.349$\times$0.272],-15.2 &11(0,11)-10(1,10) A, v$_{\rm t}$=0 & 42.52066	&-4.07280 &153.1 & 5.5\tablenotemark{c} & 5.1\tablenotemark{c} & TH + class I& Y\tablenotemark{e,f} \\
&0.0552$\times$0.049,33.3 && &  & &&\\
266838.15 (1.3E-2)&[0.256$\times$0.172],-39.0  & 5(2,4)-4(1,3) E, v$_{\rm t}$=0&	15.39064	& -4.11152	& 57.1 &9.4\tablenotemark{c} & 7.2\tablenotemark{c} & TH + class I &  Y\tablenotemark{e} \\
\enddata
\tablecomments{Line information is from CDMS. $^a$TH: Thermal. $^b$The peak flux density (F$_{peak}$) at the velocity of peak intensity ($v_{peak}$) is calculated using a 1$''$ diameter aperture centered at R.A.=6$^{h}$12$^{m}$54$^{s}$.01, decl.=17$^{d}$59$^{m}$23$^{s}$.17 (J2000) to encompass the class II maser emitting region (Figure \ref{fig_mom_classII}). $^{c}$ The same measurement (F$_{peak}$ at $v_{peak}$) is conducted with a 16$''$diameter centered at R.A.=6$^{h}$12$^{m}$53$^{s}$.64, decl.=17$^{d}$59$^{m}$22$^{s}$.77 (J2000), to cover the entire $\chthreeoh$ emission (Figure \ref{fig:chanmap_classI}).
$^d$\citet{Cragg2005} for class II, and $^e$\citet{Nesterenok2021JPhCS2103a2012N} and $^f$\citet{Voronkov2012IAUS..287..433V} for class I models, respectively.}
\end{deluxetable*}

\subsection{Class I $\chthreeoh$ maser}\label{sec:classI}
Class I $\chthreeoh$ masers are known to be collisionally pumped. Observational studies have found that they often trace the low-velocity components of the outflows along with the outflow lobes in both low- and high-mass star-forming regions \citep[e.g.,][]{Slysh1994MNRAS.268..464S,Slysh2002, Cyganowski2018IAUS..336..281C}. 

\begin{figure*}
  \centering
\includegraphics[clip,width=\textwidth,keepaspectratio]{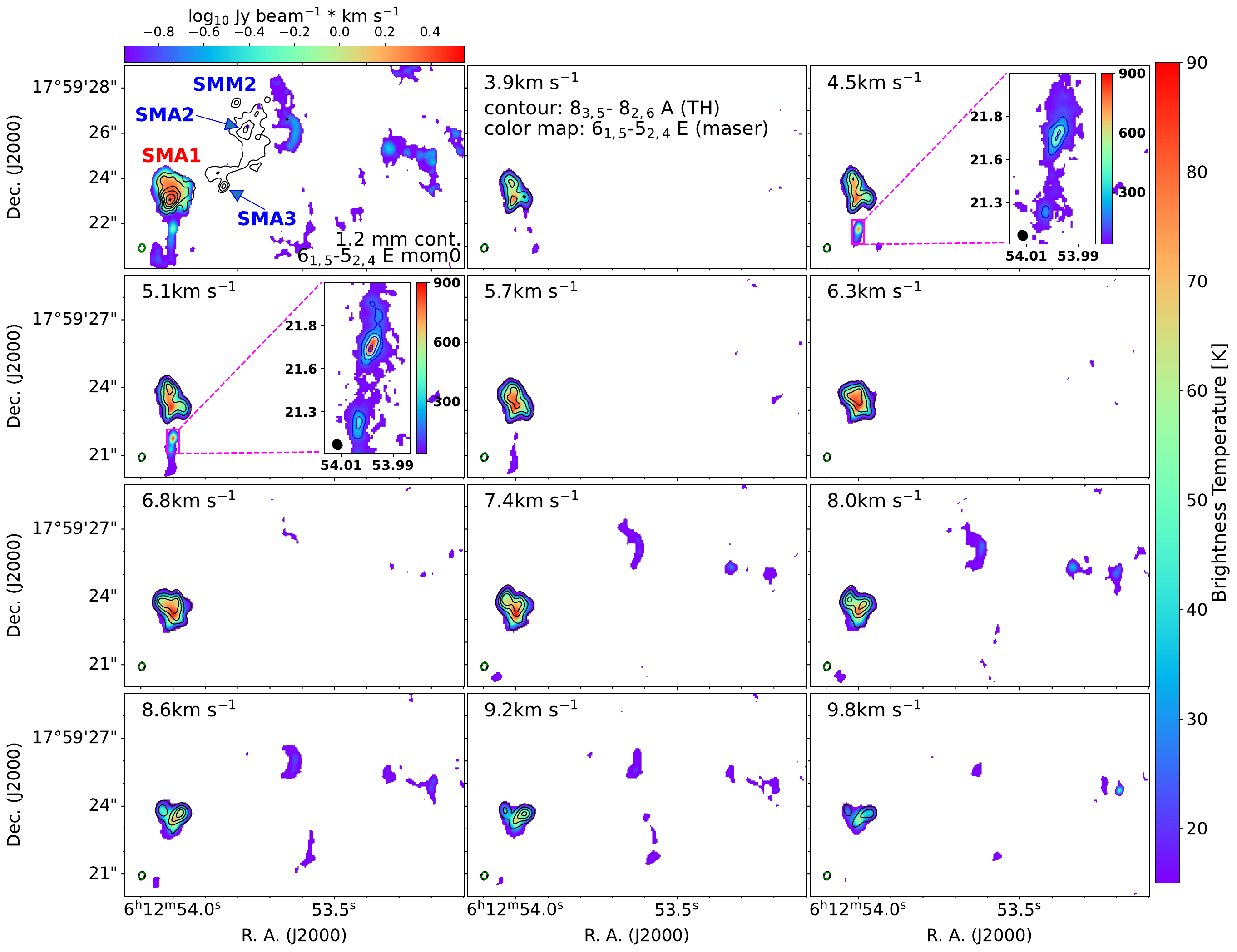}
\caption{First panel: The integrated intensity map of $\cItwofivezero$ line (color) over 1.2 mm continuum (contour). The contour levels for the continuum emission are 2.5, 4.9, 9.5, 18.6, 36.3, and 70.8 m$\jypbeam$.
Rest panels: The channel map of $\cItwofivezero$ in the unit of T$_{\rm b}$ (color) and $\thermal$ line (contour). The contour levels for the $\thermal$ emission are 15, 40, 65, and 85 K. The beam sizes are presented in the lower-left corner of each channel (Table \ref{tab:transitions}). Inset figures in 4.5 and 5.1 km s$^{-1}$ channels depict the zoom-in view of the magenta box ($\cItwofivezero$ line in both color and contour) to present the high-resolution data obtained from both compact and extended configuration combined observation. The contour levels for the $\cItwofivezero$ emission in the inset figures are 150, 300, 530, and 830 K.}
\label{fig:chanmap_classI}
\end{figure*}

Figure \ref{fig:chanmap_classI} depict the eemission distribution of $\chthreeoh$ $\cItwofivezero$ at 250506.98 MHz. The first panel presents the integrated intensity map of $\cItwofivezero$ over the 1.2 mm continuum. The continuum cores identified by previous ALMA 0.9 mm and SMA 1.3 mm observations are indicated \citep[SMA1 for S255IR NIRS3;][]{Zinchenko2015,Zinchenko2020}. The rest of the panel displays the $\cItwofivezero$ channel map (color) in the T$_{\rm b}$ unit. The channel map of one thermal line ($\thermal$) is overlaid as contours for comparison, and its integrated intensity map is shown in Figure \ref{fig_mom_classII}(a). All data is obtained by the compact configuration observation (0.33$''$$\times$0.23$''$) in addition to the compact and extended array combined data presented in the zoom-in panels (0.056$''\times$0.04$''$).

The $\cItwofivezero$ emission is also observed in the mixture of thermal and maser characteristics, but the emission distribution is spatially separated: the thermal transition ($\thermal$ in black contour) emission only appears near the protostar, tracing the inner envelope surrounding S255IR NIRS3. The $\cItwofivezero$ thermal component is coincident with $\thermal$, and additional emission components are found offset from the protostar. It consists of the filamentary structure in the south and several round and arc-shaped components with diffuse fragments scattered in the west, implying a distinctive emerging mechanism of the $\cItwofivezero$ emission offset from the protostar.

The southern components are detected in the blue-shifted to slightly red-shifted velocity channels (V$_{\rm lsr}$=5.2 km s$^{-1}$). The components are loosely connected with the W-branch of the inner envelope at 3.9 km s$^{-1}$. The brighter components appear inside the filamentary structure at 4.5--5.7 km s$^{-1}$. The inset images are the zoom-in view within the magenta box in higher spatial resolution images, indicating that the brightest feature is resolved into several emission components with the peak T$_{\rm b}$ of $\sim$900 K at 5.1 km s$^{-1}$. 
This high T$_{\rm b}$ in a narrow velocity range are the characteristics of the class I $\chthreeoh$ maser.
In more detail, the peak position of the compact and high T$_{\rm b}$ emission component is slightly moved to the south from 4.5 to 5.1 km s$^{-1}$ channels, implying that the maser components tracing different velocity are embedded in the filamentary structure.

The arc- and round-shaped emission components are detected in $\sim$6$''$ ($\sim$10000 au) and $\sim$12$''$ ($\sim$20000 au) to the west from the protostar, respectively, in the red-shifted channels (6.8--9.8 km s$^{-1}$).
The emission at $\sim$12$''$ offset from the protostar consists of relatively compact ones, indicative of the maser emission. The arc-shaped components and diffuse filamentary structure in the south might be composed of maser components in addition to the non-thermally heated emission by the outflow shock. However, it can not be ascertained with the observed T$_{\rm b}$ of a few tens of K, due to the beam dilution effect in current spatial resolution observation.

It is also worth noting that a similar emission distribution is detected in the $\cItwosixsix$ transition (Table \ref{tab:transitions}), except for the absence of the southern component. This transition was reported to show class I maser characteristics \citep{Chen2019ApJ...877...90C}.
The western arc-shape and compact emission components are more intense than those of the $\cItwofivezero$ emission: F$_{\rm peak}$ of $\cItwosixsix$ is higher than that of $\cItwofivezero$ at redshifted velocity. It could be explained by the lower energy transition of $\cItwosixsix$ than $\cItwofivezero$. With the model prediction and tentative detection of maser characteristics, we mark this transition for thermal and class I maser emission in Table \ref{tab:transitions}. 

The $\cItwofivezero$ transition is observed for the first time as a class I maser line. The intensity distribution of $\cItwofivezero$ in S255IR NIRS3 implies that various excitation mechanisms can occur in a transition at different physical conditions. 
Although the peak T$_{\rm b}$ found in southern filament fairly exceeds that in thermal transition, the actual peak T$_{\rm b}$ of maser components would be higher in better-resolution observation. To more clearly distinguish the maser components, higher angular and spectral resolution observation is required. 

\section{Discussion} \label{sec:discussion}
\subsection{The class II $\chthreeoh$ maser}
In class II $\chthreeoh$ maser, the maser emerging region is localized on the thermal emission (Figure \ref{fig:centroidmap}). The kinematic structure revealed by the maser transitions, blue-shifted acceleration to protostar from NE and SW directions, is distinctive from the thermal transitions that show continuous red- to blue-shifted velocity gradients along the NW-SE direction. 

The detected spatial extent ($\sim$1$''$) and velocity distribution of the class II $\chthreeoh$ maser lines are broadly consistent with that of $\ciizinchenko$ at 
349107.0 MHz class II $\chthreeoh$ maser (Figure 2 of \citealt{Zinchenko2017} and Figure \ref{fig_mom_classII}). In their 0.10$''$$\times$0.15$''$ resolution observation, the amplified emission was mostly found on the N-E side from the continuum peak with the blue-shifted velocity gradient. They also reported the presence of slightly red-shifted components connected to the northern direction. 

The $\ciizinchenko$ maser feature was detected in April and September 2016, with a peak flux density of 5.9 $\jypbeam$ and a flux density of 25 Jy \citep[observed around the burst peak,][]{Zinchenko2017}. The peak flux density decreased by 40\% (3.6 $\jypbeam$) in July 2017 without any noticeable structural change \citep{liu18}. In March 2019, the flux density was decayed to $\sim$5 Jy \citep{Salii2022MNRAS.512.3215S}, implying that the radiation field strength has declined in the post-burst phase.

Two class II $\chthreeoh$ maser emissions in this work were detected in 2021 (5 yrs after the burst peak, i.e., 3 yrs after the burst event ceased). 
The similar emission structure to $\ciizinchenko$ observed in 2016 raises the question of whether the physical condition for detected maser transitions has remained over 5 years, or it is just a coincidence though they trace different physical conditions. 
\citet{Salii2022MNRAS.512.3215S} suggested that the maser emission in the $J_1$-$J_0$ A$^{-+}$ series could emerge in T$_{\rm gas}$ of 110--600 K, $\chthreeoh$ specific column density (N$_{ CH_3OH}$/$\Delta$V) of 2.5$\times$10$^{8}$--5$\times$10$^{11}$ cm$^{-3}$s, and $\chthreeoh$ abundance over 10$^{-7}$. They also detected the decay of $\ciizinchenko$ and brightening of 7$_{1}$-7$_{0}$A$^{-+}$ masers in early 2019 with SMA. To reproduce the observation, the model constrained the T$_{\rm gas}$ of the masing region between 140-170 K. Meanwhile, the best-fit model for IRAM 30 m observation for $\ciizinchenko$ in a similar epoch provided T$_{\rm gas}$ of 210-240 K.

The physical condition of the inner envelope of S255IR NIRS3 triggering the maser emission could have persisted after the burst. As the radiation strength has almost returned to the quiescent phase \citep{Uchiyama2020PASJ...72....4U,Salii2022MNRAS.512.3215S}, the dust temperature of the inner envelope would be cooled down after the burst. Thus the lifetime of the maser depends on the gas temperature, gas cooling timescale, and molecular column density \citep{Cragg2005}. 
On the other hand, the $\ciizinchenko$ and class II $\chthreeoh$ masers studied in this work could have existed even in the quiescent condition before the accretion burst, showing observed flux density changes in response to variation in the radiation field.
However, the aforementioned two interpretations can be valid only if 
the masing condition of the $\ciizinchenko$ transition and transitions in this work are comparable. 
This question cannot be solved without the follow-up observation at the same transitions.
 
\citet{Moscadelli2017} and \citet{Szymczak2018A&A} compared the 6.7 GHz class II $\chthreeoh$ maser components before and after the accretion burst. Before the burst, the 6.7 GHz class II $\chthreeoh$ maser components were present nearby the protostellar position: a blue-shifted component in the N-E direction and a slight red-shifted one in the N-W direction (P position in Figure \ref{fig:centroidmap}). After the burst, both blue-shifted N-E and red-shifted N-W components appeared farther from the protostar with enhanced intensity. The variation of the red-shifted component is more robust in both intensity and emitting radii, indicated as P and A positions in Figure \ref{fig:centroidmap} \citep{Moscadelli2017,Szymczak2018A&A}.
The distribution is different between 6.7 GHz and millimeter class II $\chthreeoh$ masers: while 6.7 GHz maser is more prominent in red-shifted components right after the burst (observed in 2016), the millimeter wavelengths maser emissions are stronger in blue-shifted components, both in 349107.0 MHz (in 2016) and our work (in 2021). If the red-shifted maser components are present in our observation, the maser feature would be distinct even though the emission is mixed with the thermal emissions. Thus the weaker red-shifted components observed in this study and \citet{Zinchenko2017} are not the observational effect. 

The weaker intensity distribution in millimeter masers compared to 6.7 GHz masers can be caused by both the different emerging conditions and the time variation. According to \citet{Cragg2005}, the 6.7 GHz class II $\chthreeoh$ maser occurs in T$_{\rm dust}>\sim$150 K and n$_{\rm gas}<$ 10$^{8}$ cm$^{-3}$. However, the physical condition of the millimeter maser has not been constrained yet in detail. For comparing two transition observations, further observations at the same epoch and a detailed $\chthreeoh$ maser model extending to shorter wavelengths and weaker transitions are required.

\subsection{The class I $\chthreeoh$ maser}
The class I $\chthreeoh$ maser feature in the southern filamentary structure (Figure \ref{fig:chanmap_classI}) is observed near the systemic velocity (4.5 -- 5.7 $\kms$), and the western arc-shaped emission in our $\chthreeoh$ observation is red-shifted (7.5 -- 9.8 $\kms$). The velocity and emission distributions coincide with those of C$^{34}$S line \citep{Zinchenko2020}, and are inconsistent with the CO emission. It implies that both C$^{34}$S and class I $\chthreeoh$ maser are excited by the low-velocity shock, whereas CO is abundant in gas-phase by various reactions at a much wider range of velocity components ($\sim$ -70 -- 90 $\kms$).

\section{Conclusion} \label{sec:conclusion}
Even after the burst ceased, 6 additional $\chthreeoh$ transitions were found with maser characteristics in S255IR NIRS3. We detect four class II (including candidates) and two class I $\chthreeoh$ maser transitions, in the post-burst phase of S255IR NIRS3. The $\cIItwosixfive$ and $\cItwofivezero$ lines are detected as maser lines for the first time.
All transitions show a mixture of maser and thermal emission characteristics, but the observed maser features are distinctive from LTE analysis results and thermal emission features. 
In the lack of the observed data at the same epoch as well as theoretical study, the detected maser features cannot be directly compared with previous observations. To correctly understand the maser characteristics under the condition of S255IR NIRS3, monitoring observation is essential.

\begin{acknowledgments}
This research was supported by Basic Science Research Program through the National Research Foundation of Korea(NRF) funded by the Ministry of Education(RS-2023-00247790). This work was also supported by the New Faculty Startup Fund from Seoul National University and the NRF grant funded by the Korean government (MSIT) (grant number 2021R1A2C1011718).
This paper makes use of the following ALMA data: ADS/JAO.ALMA\#2019.1.00571.S, ADS/JAO.ALMA\#2021.1.01056.S. ALMA is a partnership of ESO (representing its member states), NSF (USA) and NINS (Japan), together with NRC (Canada), MOST and ASIAA (Taiwan), and KASI (Republic of Korea), in cooperation with the Republic of Chile. The Joint ALMA Observatory is operated by ESO, AUI/NRAO and NAOJ. 
\end{acknowledgments}

\facilities{Atacama Large Millimeter/submillimeter Array
(ALMA).}
\software{Common Astronomy Software Applications package (CASA; \citealt{CASA2022PASP..134k4501C}), 
STATCONT \citep{SanchezMonge2018},
eXtended CASA Line Analysis Software Suite (XCLASS; \citealt{Moller2017})
          }

\bibliographystyle{aasjournal}
\bibliography{ms}{}

\begin{thebibliography}{}
\expandafter\ifx\csname natexlab\endcsname\relax\def\natexlab#1{#1}\fi
\providecommand{\url}[1]{\href{#1}{#1}}
\providecommand{\dodoi}[1]{doi:~\href{http://doi.org/#1}{\nolinkurl{#1}}}
\providecommand{\doeprint}[1]{\href{http://ascl.net/#1}{\nolinkurl{http://ascl.net/#1}}}
\providecommand{\doarXiv}[1]{\href{https://arxiv.org/abs/#1}{\nolinkurl{https://arxiv.org/abs/#1}}}

\bibitem[{{Baek} {et~al.}(2022){Baek}, {Lee}, {Hirota}, {Kim}, \&
  {Kim}}]{Baek2022ApJ...939...84B}
{Baek}, G., {Lee}, J.-E., {Hirota}, T., {Kim}, K.-T., \& {Kim}, M.~K. 2022,
  \apj, 939, 84, \dodoi{10.3847/1538-4357/ac81d3}

\bibitem[{{Brogan} {et~al.}(2019){Brogan}, {Hunter}, {Towner}, {McGuire},
  {MacLeod}, {Gurwell}, {Cyganowski}, {Brand}, {Burns}, {Caratti o Garatti},
  {Chen}, {Chibueze}, {Hirano}, {Hirota}, {Kim}, {Kramer}, {Linz}, {Menten},
  {Remijan}, {Sanna}, {Sobolev}, {Sridharan}, {Stecklum}, {Sugiyama}, {Surcis},
  {Van der Walt}, {Volvach}, \& {Volvach}}]{Brogan2019ApJ}
{Brogan}, C.~L., {Hunter}, T.~R., {Towner}, A.~P.~M., {et~al.} 2019, \apjl,
  881, L39, \dodoi{10.3847/2041-8213/ab2f8a}

\bibitem[{{Burns} {et~al.}(2016){Burns}, {Handa}, {Nagayama}, {Sunada}, \&
  {Omodaka}}]{Burns2016}
{Burns}, R.~A., {Handa}, T., {Nagayama}, T., {Sunada}, K., \& {Omodaka}, T.
  2016, \mnras, 460, 283, \dodoi{10.1093/mnras/stw958}

\bibitem[{{Burns} {et~al.}(2020){Burns}, {Sugiyama}, {Hirota}, {Kim},
  {Sobolev}, {Stecklum}, {MacLeod}, {Yonekura}, {Olech}, {Orosz}, {Ellingsen},
  {Hyland}, {Caratti o Garatti}, {Brogan}, {Hunter}, {Phillips}, {van den
  Heever}, {Eisl{\"o}ffel}, {Linz}, {Surcis}, {Chibueze}, {Baan}, \&
  {Kramer}}]{Burns2020NatAs...4..506B}
{Burns}, R.~A., {Sugiyama}, K., {Hirota}, T., {et~al.} 2020, Nature Astronomy,
  4, 506, \dodoi{10.1038/s41550-019-0989-3}

\bibitem[{{Burns} {et~al.}(2023){Burns}, {Uno}, {Sakai}, {Blanchard}, {Rosli},
  {Orosz}, {Yonekura}, {Tanabe}, {Sugiyama}, {Hirota}, {Kim}, {Aberfelds},
  {Volvach}, {Bartkiewicz}, {Caratti o Garatti}, {Sobolev}, {Stecklum},
  {Brogan}, {Phillips}, {Ladeyschikov}, {Johnstone}, {Surcis}, {MacLeod},
  {Linz}, {Chibueze}, {Brand}, {Eisl{\"o}ffel}, {Hyland}, {Uscanga}, {Olech},
  {Durjasz}, {Bayandina}, {Breen}, {Ellingsen}, {van den Heever}, {Hunter}, \&
  {Chen}}]{Burns2023NatAs.tmp...57B}
{Burns}, R.~A., {Uno}, Y., {Sakai}, N., {et~al.} 2023, Nature Astronomy,
  \dodoi{10.1038/s41550-023-01899-w}

\bibitem[{{Caratti o Garatti} {et~al.}(2017){Caratti o Garatti}, {Stecklum},
  {Garcia Lopez}, {Eisl{\"o}ffel}, {Ray}, {Sanna}, {Cesaroni}, {Walmsley},
  {Oudmaijer}, {de Wit}, {Moscadelli}, {Greiner}, {Krabbe}, {Fischer}, {Klein},
  \& {Iba{\~n}ez}}]{CarattioGaratti2017}
{Caratti o Garatti}, A., {Stecklum}, B., {Garcia Lopez}, R., {et~al.} 2017,
  Nature Physics, 13, 276, \dodoi{10.1038/nphys3942}

\bibitem[{{CASA Team} {et~al.}(2022){CASA Team}, {Bean}, {Bhatnagar}, {Castro},
  {Donovan Meyer}, {Emonts}, {Garcia}, {Garwood}, {Golap}, {Gonzalez Villalba},
  {Harris}, {Hayashi}, {Hoskins}, {Hsieh}, {Jagannathan}, {Kawasaki},
  {Keimpema}, {Kettenis}, {Lopez}, {Marvil}, {Masters}, {McNichols},
  {Mehringer}, {Miel}, {Moellenbrock}, {Montesino}, {Nakazato}, {Ott}, {Petry},
  {Pokorny}, {Raba}, {Rau}, {Schiebel}, {Schweighart}, {Sekhar}, {Shimada},
  {Small}, {Steeb}, {Sugimoto}, {Suoranta}, {Tsutsumi}, {van Bemmel},
  {Verkouter}, {Wells}, {Xiong}, {Szomoru}, {Griffith}, {Glendenning}, \&
  {Kern}}]{CASA2022PASP..134k4501C}
{CASA Team}, {Bean}, B., {Bhatnagar}, S., {et~al.} 2022, \pasp, 134, 114501,
  \dodoi{10.1088/1538-3873/ac9642}

\bibitem[{{Cesaroni} {et~al.}(2018){Cesaroni}, {Moscadelli}, {Neri}, {Sanna},
  {Caratti o Garatti}, {Eisloffel}, {Stecklum}, {Ray}, \&
  {Walmsley}}]{Cesaroni2018}
{Cesaroni}, R., {Moscadelli}, L., {Neri}, R., {et~al.} 2018, \aap, 612, A103,
  \dodoi{10.1051/0004-6361/201732238}

\bibitem[{{Chen} {et~al.}(2019){Chen}, {Ellingsen}, {Ren}, {Sobolev},
  {Parfenov}, \& {Shen}}]{Chen2019ApJ...877...90C}
{Chen}, X., {Ellingsen}, S.~P., {Ren}, Z.-Y., {et~al.} 2019, \apj, 877, 90,
  \dodoi{10.3847/1538-4357/ab1078}

\bibitem[{{Chen} {et~al.}(2020){Chen}, {Sobolev}, {Ren}, {Parfenov}, {Breen},
  {Ellingsen}, {Shen}, {Li}, {MacLeod}, {Baan}, {Brogan}, {Hirota}, {Hunter},
  {Linz}, {Menten}, {Sugiyama}, {Stecklum}, {Gong}, \&
  {Zheng}}]{Chen2020NatAs...4.1170C}
{Chen}, X., {Sobolev}, A.~M., {Ren}, Z.-Y., {et~al.} 2020, Nature Astronomy, 4,
  1170, \dodoi{10.1038/s41550-020-1144-x}

\bibitem[{{Cragg} {et~al.}(2005){Cragg}, {Sobolev}, \& {Godfrey}}]{Cragg2005}
{Cragg}, D.~M., {Sobolev}, A.~M., \& {Godfrey}, P.~D. 2005, \mnras, 360, 533,
  \dodoi{10.1111/j.1365-2966.2005.09077.x}

\bibitem[{{Cyganowski} {et~al.}(2018){Cyganowski}, {Hannaway}, {Brogan},
  {Hunter}, \& {Zhang}}]{Cyganowski2018IAUS..336..281C}
{Cyganowski}, C.~J., {Hannaway}, D., {Brogan}, C.~L., {Hunter}, T.~R., \&
  {Zhang}, Q. 2018, in Astrophysical Masers: Unlocking the Mysteries of the
  Universe, ed. A.~{Tarchi}, M.~J. {Reid}, \& P.~{Castangia}, Vol. 336,
  281--282, \dodoi{10.1017/S1743921317010717}

\bibitem[{{Fujisawa} {et~al.}(2015){Fujisawa}, {Yonekura}, {Sugiyama},
  {Horiuchi}, {Hayashi}, {Hachisuka}, {Matsumoto}, \&
  {Niinuma}}]{Fujisawa2015ATel.8286....1F}
{Fujisawa}, K., {Yonekura}, Y., {Sugiyama}, K., {et~al.} 2015, The Astronomer's
  Telegram, 8286, 1

\bibitem[{{Goddi} {et~al.}(2007){Goddi}, {Moscadelli}, {Sanna}, {Cesaroni}, \&
  {Minier}}]{Goddi2007}
{Goddi}, C., {Moscadelli}, L., {Sanna}, A., {Cesaroni}, R., \& {Minier}, V.
  2007, \aap, 461, 1027, \dodoi{10.1051/0004-6361:20066136}

\bibitem[{Hern{\'{a}}ndez-Hern{\'{a}}ndez
  {et~al.}(2014)Hern{\'{a}}ndez-Hern{\'{a}}ndez, Zapata, Kurtz, \&
  Garay}]{Hernandez-Hernandez2014}
Hern{\'{a}}ndez-Hern{\'{a}}ndez, V., Zapata, L., Kurtz, S., \& Garay, G. 2014,
  The Astrophysical Journal, 786, 38, \dodoi{10.1088/0004-637X/786/1/38}

\bibitem[{{Hirota} {et~al.}(2021){Hirota}, {Cesaroni}, {Moscadelli},
  {Sugiyama}, {Burns}, {Kim}, {Sunada}, \& {Yonekura}}]{Hirota2021}
{Hirota}, T., {Cesaroni}, R., {Moscadelli}, L., {et~al.} 2021, \aap, 647, A23,
  \dodoi{10.1051/0004-6361/202039798}

\bibitem[{{Hirota} {et~al.}(2022){Hirota}, {Wolak}, {Hunter}, {Brogan},
  {Bartkiewicz}, {Durjasz}, {Kobak}, {Olech}, {Szymczak}, {Burns}, {Aberfelds},
  {Baek}, {Brand}, {Breen}, {Byun}, {Caratti o Garatti}, {Chen}, {Chibueze},
  {Cyganowski}, {Eisl{\"o}ffel}, {Ellingsen}, {Hirano}, {Hu}, {Kang}, {Kim},
  {Kim}, {Kim}, {Kim}, {Kramer}, {Lee}, {Linz}, {Liu}, {MacLeod}, {McCarthy},
  {Menten}, {Motogi}, {Oh}, {Orosz}, {Sobolev}, {Stecklum}, {Sugiyama},
  {Sunada}, {Uscanga}, {van den Heever}, {Volvach}, {Volvach}, {Wu}, \&
  {Yonekura}}]{Hirota2022PASJ...74.1234H}
{Hirota}, T., {Wolak}, P., {Hunter}, T.~R., {et~al.} 2022, \pasj, 74, 1234,
  \dodoi{10.1093/pasj/psac067}

\bibitem[{{Humire} {et~al.}(2022){Humire}, {Henkel}, {Hern{\'a}ndez-G{\'o}mez},
  {Mart{\'\i}n}, {Mangum}, {Harada}, {Muller}, {Sakamoto}, {Tanaka},
  {Yoshimura}, {Nakanishi}, {M{\"u}hle}, {Herrero-Illana}, {Meier}, {Caux},
  {Aladro}, {Mauersberger}, {Viti}, {Colzi}, {Rivilla}, {Gorski}, {Menten},
  {Huang}, {Aalto}, {van der Werf}, \& {Emig}}]{Humire2022A&A...663A..33H}
{Humire}, P.~K., {Henkel}, C., {Hern{\'a}ndez-G{\'o}mez}, A., {et~al.} 2022,
  \aap, 663, A33, \dodoi{10.1051/0004-6361/202243384}

\bibitem[{{Hunter} {et~al.}(2017){Hunter}, {Brogan}, {MacLeod}, {Cyganowski},
  {Chandler}, {Chibueze}, {Friesen}, {Indebetouw}, {Thesner}, \&
  {Young}}]{hunter17}
{Hunter}, T.~R., {Brogan}, C.~L., {MacLeod}, G., {et~al.} 2017, \apjl, 837,
  L29, \dodoi{10.3847/2041-8213/aa5d0e}

\bibitem[{{Kim} {et~al.}(2020){Kim}, {Kim}, {Hirota}, {Kim}, {Sugiyama},
  {Honma}, {Byun}, {Oh}, {Motogi}, {Kang}, {Kim}, {Liu}, {Hu}, {Burns},
  {Chibueze}, {Matsumoto}, \& {Sunada}}]{Kim2020}
{Kim}, J., {Kim}, M.~K., {Hirota}, T., {et~al.} 2020, \apj, 896, 127,
  \dodoi{10.3847/1538-4357/ab9100}

\bibitem[{{Liu} {et~al.}(2020){Liu}, {Su}, {Zinchenko}, {Wang}, {Meyer},
  {Wang}, \& {Hsieh}}]{Liu2020}
{Liu}, S.-Y., {Su}, Y.-N., {Zinchenko}, I., {et~al.} 2020, \apj, 904, 181,
  \dodoi{10.3847/1538-4357/abc0ec}

\bibitem[{{Liu} {et~al.}(2018){Liu}, {Su}, {Zinchenko}, {Wang}, \&
  {Wang}}]{liu18}
{Liu}, S.-Y., {Su}, Y.-N., {Zinchenko}, I., {Wang}, K.-S., \& {Wang}, Y. 2018,
  The Astrophysical Journal, 863, L12, \dodoi{10.3847/2041-8213/aad63a}

\bibitem[{{MacLeod} {et~al.}(2019){MacLeod}, {Sugiyama}, {Hunter}, {Quick},
  {Baan}, {Breen}, {Brogan}, {Burns}, {Caratti o Garatti}, {Chen}, {Chibueze},
  {Houde}, {Kaczmarek}, {Linz}, {Rajabi}, {Saito}, {Schmidl}, {Sobolev},
  {Stecklum}, {van den Heever}, \& {Yonekura}}]{MacLeod2019MNRAS.489.3981M}
{MacLeod}, G.~C., {Sugiyama}, K., {Hunter}, T.~R., {et~al.} 2019, \mnras, 489,
  3981, \dodoi{10.1093/mnras/stz2417}

\bibitem[{{Meyer} {et~al.}(2019){Meyer}, {Vorobyov}, {Elbakyan}, {Stecklum},
  {Eisl{\"o}ffel}, \& {Sobolev}}]{Meyer2019MNRAS.482.5459M}
{Meyer}, D.~M.~A., {Vorobyov}, E.~I., {Elbakyan}, V.~G., {et~al.} 2019, \mnras,
  482, 5459, \dodoi{10.1093/mnras/sty2980}

\bibitem[{{Meyer} {et~al.}(2017){Meyer}, {Vorobyov}, {Kuiper}, \&
  {Kley}}]{Meyer2017MNRAS.464L..90M}
{Meyer}, D.~M.~A., {Vorobyov}, E.~I., {Kuiper}, R., \& {Kley}, W. 2017, \mnras,
  464, L90, \dodoi{10.1093/mnrasl/slw187}

\bibitem[{{M{\"o}ller} {et~al.}(2017){M{\"o}ller}, {Endres}, \&
  {Schilke}}]{Moller2017}
{M{\"o}ller}, T., {Endres}, C., \& {Schilke}, P. 2017, \aap, 598, A7,
  \dodoi{10.1051/0004-6361/201527203}

\bibitem[{{Moscadelli} {et~al.}(2017){Moscadelli}, {Sanna}, {Goddi},
  {Walmsley}, {Cesaroni}, {Caratti o Garatti}, {Stecklum}, {Menten}, \&
  {Kraus}}]{Moscadelli2017}
{Moscadelli}, L., {Sanna}, A., {Goddi}, C., {et~al.} 2017, \aap, 600, L8,
  \dodoi{10.1051/0004-6361/201730659}

\bibitem[{{Nesterenok}(2021)}]{Nesterenok2021JPhCS2103a2012N}
{Nesterenok}, A.~V. 2021, in Journal of Physics Conference Series, Vol. 2103,
  Journal of Physics Conference Series, 012012,
  \dodoi{10.1088/1742-6596/2103/1/012012}

\bibitem[{{Salii} {et~al.}(2022){Salii}, {Zinchenko}, {Liu}, {Sobolev},
  {Aberfelds}, \& {Su}}]{Salii2022MNRAS.512.3215S}
{Salii}, S.~V., {Zinchenko}, I.~I., {Liu}, S.-Y., {et~al.} 2022, \mnras, 512,
  3215, \dodoi{10.1093/mnras/stac739}

\bibitem[{{S{\'a}nchez-Monge} {et~al.}(2018){S{\'a}nchez-Monge}, {Schilke},
  {Ginsburg}, {Cesaroni}, \& {Schmiedeke}}]{SanchezMonge2018}
{S{\'a}nchez-Monge}, {\'A}., {Schilke}, P., {Ginsburg}, A., {Cesaroni}, R., \&
  {Schmiedeke}, A. 2018, \aap, 609, A101, \dodoi{10.1051/0004-6361/201730425}

\bibitem[{{Slysh} {et~al.}(1994){Slysh}, {Kalenskii}, {Valtts}, \&
  {Otrupcek}}]{Slysh1994MNRAS.268..464S}
{Slysh}, V.~I., {Kalenskii}, S.~V., {Valtts}, I.~E., \& {Otrupcek}, R. 1994,
  \mnras, 268, 464, \dodoi{10.1093/mnras/268.2.464}

\bibitem[{{Slysh} {et~al.}(2002){Slysh}, {Kalenski{\u{i}}}, \&
  {Val'tts}}]{Slysh2002}
{Slysh}, V.~I., {Kalenski{\u{i}}}, S.~V., \& {Val'tts}, I.~E. 2002, Astronomy
  Reports, 46, 49, \dodoi{10.1134/1.1436204}

\bibitem[{{Stecklum} {et~al.}(2016){Stecklum}, {Caratti o Garatti}, {Cardenas},
  {Greiner}, {Kruehler}, {Klose}, \&
  {Eisloeffel}}]{Stecklum2016ATel.8732....1S}
{Stecklum}, B., {Caratti o Garatti}, A., {Cardenas}, M.~C., {et~al.} 2016, The
  Astronomer's Telegram, 8732, 1

\bibitem[{{Szymczak} {et~al.}(2018){Szymczak}, {Olech}, {Wolak}, {G{\'e}rard},
  \& {Bartkiewicz}}]{Szymczak2018A&A}
{Szymczak}, M., {Olech}, M., {Wolak}, P., {G{\'e}rard}, E., \& {Bartkiewicz},
  A. 2018, \aap, 617, A80, \dodoi{10.1051/0004-6361/201833443}

\bibitem[{{Uchiyama} {et~al.}(2020){Uchiyama}, {Yamashita}, {Sugiyama},
  {Nakaoka}, {Kawabata}, {Itoh}, {Yamanaka}, {Akitaya}, {Kawabata}, {Yonekura},
  {Saito}, {Motogi}, \& {Fujisawa}}]{Uchiyama2020PASJ...72....4U}
{Uchiyama}, M., {Yamashita}, T., {Sugiyama}, K., {et~al.} 2020, \pasj, 72, 4,
  \dodoi{10.1093/pasj/psz122}

\bibitem[{{Volvach} {et~al.}(2020){Volvach}, {Volvach}, {Larionov}, {MacLeod},
  {van den Heever}, \& {Sugiyama}}]{Volvach2020MNRAS.494L..59V}
{Volvach}, A.~E., {Volvach}, L.~N., {Larionov}, M.~G., {et~al.} 2020, \mnras,
  494, L59, \dodoi{10.1093/mnrasl/slaa036}

\bibitem[{{Voronkov} {et~al.}(2012){Voronkov}, {Caswell}, {Ellingsen}, {Breen},
  {Britton}, {Green}, {Sobolev}, \& {Walsh}}]{Voronkov2012IAUS..287..433V}
{Voronkov}, M.~A., {Caswell}, J.~L., {Ellingsen}, S.~P., {et~al.} 2012, in
  Cosmic Masers - from OH to H0, ed. R.~S. {Booth}, W.~H.~T. {Vlemmings}, \&
  E.~M.~L. {Humphreys}, Vol. 287, 433--440, \dodoi{10.1017/S174392131200748X}

\bibitem[{{Zinchenko} {et~al.}(2017){Zinchenko}, {Liu}, {Su}, \&
  {Sobolev}}]{Zinchenko2017}
{Zinchenko}, I., {Liu}, S.~Y., {Su}, Y.~N., \& {Sobolev}, A.~M. 2017, \aap,
  606, L6, \dodoi{10.1051/0004-6361/201731752}

\bibitem[{{Zinchenko} {et~al.}(2015){Zinchenko}, {Liu}, {Su}, {Salii},
  {Sobolev}, {Zemlyanukha}, {Beuther}, {Ojha}, {Samal}, \&
  {Wang}}]{Zinchenko2015}
{Zinchenko}, I., {Liu}, S.~Y., {Su}, Y.~N., {et~al.} 2015, \apj, 810, 10,
  \dodoi{10.1088/0004-637X/810/1/10}

\bibitem[{{Zinchenko} {et~al.}(2020){Zinchenko}, {Liu}, {Su}, {Wang}, \&
  {Wang}}]{Zinchenko2020}
{Zinchenko}, I.~I., {Liu}, S.-Y., {Su}, Y.-N., {Wang}, K.-S., \& {Wang}, Y.
  2020, \apj, 889, 43, \dodoi{10.3847/1538-4357/ab5c18}

\end{thebibliography}

\end{document}